\documentclass{nature}
\usepackage{amsmath}
\usepackage{amsfonts}
\usepackage{amssymb}
\usepackage{graphicx}
\usepackage{dcolumn}
\usepackage{bm}
\usepackage{color}

\title{Topological States on the Gold Surface} 
\author
{Binghai Yan$^{1,2~\ast}$, Benjamin Stadtm{\"u}ller$^{3}$, Norman Haag$^{3}$, Sebastian Jakobs$^{3}$, Johannes Seidel$^{3}$,  Dominik Jungkenn$^{3}$,  Stefan Mathias$^{3}$, Mirko Cinchetti$^{3}$, Martin Aeschlimann$^{3}$ and Claudia Felser$^{1}$}
\date{\today}
\begin{document}
\maketitle

\begin{affiliations}
\item Max Planck Institute for Chemical Physics of Solids, 01187 Dresden, Germany
\item Max Planck Institute for the Physics of Complex Systems, 01187 Dresden, Germany
\item Department of Physics and Research Center OPTIMAS, University of Kaiserslautern, 67653 Kaiserslautern, Germany
\end{affiliations}


\begin{abstract}
 Gold surfaces host special electronic states~\cite{Heimann1977,LaShell1996}
that have been understood as a prototype of Shockley surface states (SSs)~\cite{Shockley1939}.
These SSs are commonly employed to benchmark the capability of angle-resolved photoemission spectroscopy (ARPES) ~\cite{Kliewer2000,Reinert2003} and scanning tunneling spectroscopy ~\cite{Burgi1999}.
 We find that these Shockley SSs can be reinterpreted as {\color{black}topologically derived surface states (TDSSs)} of a topological insulator (TI)~\cite{qi2010,moore2010,hasan2010,qi2011RMP}, a recently discovered quantum state. 
Based on band structure calculations, the $Z_2$ topological invariant  can be well defined to characterize the nontrivial features of gold that we detect by ARPES. The same TDSSs are also recognized on surfaces of other well-known noble  metals (e.g., silver, copper, platinum, and palladium).  
Besides  providing a new understanding of noble metal SSs,
finding topological states on late transition metals provokes  interesting questions on the role of topological effects in surface-related processes, such as adsorption and catalysis.
\end{abstract}

The history of surface states (SSs) can be traced back to 1932 when Tamm~\cite{Tamm1932} predicted the existence of special electronic states near the crystal boundary. Soon Shockley~\cite{Shockley1939} found that SSs, usually called Shockley SSs later, emerge in an inverted energy gap owing to band crossing, for which the symmetry of the bulk band structure was found to be crucial~\cite{Zak1985}. The SSs on (111) surfaces of noble metals (e.g., Au, Ag, and Cu) have been known as typical Shockley-type SSs (e.g., refs.~\cite{Gartland1975,Kevan1987,Reinert2001}), wherein SSs appear inside an inverted energy gap of $s$ and $p$ bands at the center of the surface Brillouin zone. As a result of  inversion symmetry breaking on the surface, these SSs exhibit Rashba-type~\cite{Rashba1984} spin splitting with spin-momentum locking at the Fermi surface~\cite{LaShell1996,Reinert2001,Hoesch2004,Tamai2013}, 
which is essential for spintronic devices~\cite{Wolf2001}. These SSs have been used to
design quantum corrals~\cite{Crommie1993,Heller1994,Burgi1998,Mugarza2001} as well as artificial Dirac fermions~\cite{Gomes2012} and benchmark the capability of angle-resolved photoemission spectroscopy (ARPES) ~\cite{Kliewer2000,Reinert2003} and scanning tunneling spectroscopy (STS) ~\cite{Burgi1999}.
As another intrinsic SS, the topological surface state (TSS) has recently  attracted great research interest in the condensed matter physics community ~\cite{hasan2010,qi2011RMP}. Metallic TSSs inside the bulk energy gap are induced by the topology of the inherent bulk band structure, which can be understood as an inversion between the conduction and valence bands that have opposite parities~\cite{bernevig2006d,fu2007a}. TSSs have been  predicted and observed  in many compounds~\cite{Yan2012rpp}, such as HgTe~\cite{bernevig2006d,koenig2007} and Bi$_2$Se$_3$ ~\cite{zhang2009,xia2009}, wherein spin and momentum are locked up and form spin texture in the Dirac-cone-like band structure. 
From a naive viewpoint of the band inversion, Shockley SSs and TSSs are not fully exclusive of each other.  
TSSs of Bi$_2$Se$_3$, for example,  has been described in a generalized Shockley model that includes spin-orbit coupling (SOC) by Pershoguba \textit{et al.}~\cite{Pershoguba2012}. This inspires us to pose the opposite question: Can some Shockley SSs be understood as TSSs? 

In this Letter, we revisit SSs on Au, Ag, and Cu (111) surfaces by \textit{ab initio} band structure calculations and angle-resolved photoemission spectroscopy (ARPES) performed on the Au(111) surface with a {\it momentum microscope} \cite{kspace} for the detection of the complete angular distribution of the photoemitted electrons as function of their kinetic energy. We find that these famous Shockley SSs are also TSSs, which originate from the inverted bulk band structure. The Rashba-split-like energy dispersion can be regarded as a strongly distorted Dirac cone. Although noble metals do not exhibit an energy gap, a $Z_2$ topological invariant $\nu_0=1$ can be well defined owing to the existence of a direct energy gap above the Fermi energy. The existence of topological states in real materials is more ubiquitous than ever believed.


Noble metals  share the same face-centered-cubic (FCC) lattice structure. We take Au as an example. As shown in Fig. 1a, the primitive unit cell includes a single Au atom as the inversion center of the lattice. Therefore, the parity of the Bloch wave function is consistent with that of the corresponding atomic orbital: ``+'' for the Au-$s$ and $d$ orbitals and ``$-$'' for the Au-$p$ orbital. The bulk band structure that includes the SOC effect is shown in Fig. 1c. 
The Fermi energy crosses the middle of a band wherein $s$ (blue color) and $p$ (red color) states hybridize together, wherein the $d$ bands are fully occupied and below the Fermi energy. 
Because the band inversion only involves $sp$ rather than $d$ states, 
we project the band structure to Au-$sp$ states using Wannier function method for simplicity.
Above the half-filled band, a direct energy gap exists in the whole Brillouin zone owing to SOC, as indicated by the gray shadow in Fig. 1c, though the indirect gap is still zero.  As we will see, this direct energy gap determines the topology of SSs. For the sake of simplicity, we call bands below and above the gap $valence$ and $conduction$ bands, respectively.  
To illustrate the band inversion clearly, we show the energy dispersion along $\Gamma-X^{\prime}-L^{\prime}-\Gamma$ lines, wherein $X^{\prime}$($L^{\prime}$) is equivalent to $X$ ($L$).
Because of relativistic contraction of Au-$6s$ orbitals~\cite[and references therein]{Pyykko2013}, 
the $s$ band is lower in energy than the $d$ and $p$ bands at the $\Gamma$ point. 
In contrast, the $s$ band energy is even higher than that of $p$ bands at $X^{\prime}$ ($X$) and $L^{\prime}$ ($L$) points. Thus,  one can find that $s$ and $p$ bands preserve the normal order at the $\Gamma$ point but get inverted at other time-reversal-invariant-momenta (TRIM) $X^{\prime}$ ($X$) and $L^{\prime}$ ($L$) points.
Between the $\Gamma$ and $X^{\prime}$ ($L^{\prime}$) points the $s$ and $p$ bands cross each other. The direct energy gap opens at the crossing point because of band anticrossing caused by SOC.
The $Z_2$ topological invariant $\nu_0$ of a topological insulator (TI)  can be calculated by the product of the parity eigenvalues of all valence bands at all TRIM~\cite{fu2007a}. If the parity product is $-1=(-1)^{\nu_0}$, then $\nu_0=1$ represents a TI; otherwise, $\nu_0=0$ represents a trivial insulator.
In the FCC Brillouin zone, eight TRIM include one $\Gamma$ point, three $X$ points, and four $L$ points. Because $s$ and $d$ states are always ``$+$'' in parity while only $p$ states are ``$-$'', the parity product at a given $k$ point is determined by the number of $p$ states. Therefore, the parity product of the $\Gamma$ point is ``+'' since only $sd$ states appear in the valence bands. However, the parity products at $X$ and $L$ are ``$-$'', for there is one $p$ state as the top valence band for both $X$ and $L$ points. Thus, the total parity product for eight TRIM is ``$-$'', i.e., the $Z_2$ invariant $\nu_0=1$, showing the topologically nontrivial feature. One can see that the topology of the band structure is caused by the $s$--$p$ band inversion above the Fermi energy, which is related to the relativistic contraction of the Au-$6s$ state.

The nontrivial $Z_2$ topological number guarantees the existence of TSSs on the boundary. On the  Au(111) surface, the $s$--$p$ inversion gap (also called $L$-gap in the literature) remains at the $\bar{\Gamma}$ point of the surface Brillouin zone while it is reduced to zero at $\bar{M}$ and $\bar{K}$ points. Near the $\bar{\Gamma}$ point, a pair of TSSs exist inside the  $s$--$p$ inversion gap with the Dirac point lying below the Fermi energy, as shown in Fig. 2a.
Although the topological energy gap is above the Fermi energy, the local surface potential pulls the Dirac cone below the Fermi energy (see Supplementary Information). 
As shown in Fig. 2a, the Dirac cone is strongly distorted, with the left-hand spin texture in the upper cone being similar to known TIs~\cite{hsieh2009,liu2010}.
Analysis of orbital components reveals that TSSs are mainly composed by $sp$ orbitals. 
 The same energy dispersion has been previous observed using ARPES and revealed in \textit{ab-initio} calculations~\cite{LaShell1996,Hoesch2004,Reinert2001,Henk2003}; 
this dispersion was interpreted as a Rashba-type split of $sp$-derived Shockley SSs. 
However,  in a local region (e.g., near the $\bar{\Gamma}$ point) in the Brillouin zone, one cannot distinguish TSSs from trivial Rashba SSs. These Rashba states are equivalent to TSSs if the Rashba-split bands are regard as a strongly distorted Dirac cone wherein the lower cone was pushed above the Dirac point. 
A similar dispersion of TSSs as a dramatically deformed Dirac cone was recently observed on the surface of HgTe~\cite{hancock2011}.

To prove the topological origin of these SSs, we follow a twofold strategy. First of all, we use two-photon-photoemission (2PPE) ARPES  to measure the energy dispersion of the SSs of the Au(111) surface below and above the Fermi energy. 
By combining an optical parametric oscillator (OPO) laser system with a modern momentum microscope \cite{kspace} we are able to map for the first time the dispersion of the SSs far away from the $\bar{\Gamma}$ point, and confirm experimentally the strong deviation from the dispersion of a trivial Rashba SSs. Second, we demonstrate theoretically that these SSs are adiabatically connected to TSSs of a real TI.  \\
We first describe the ARPES results. In order to detect the electronic structure of the Au(111) surface both below and above the Fermi energy ($E_F$), we have performed 2PPE ARPES with an OPO laser system and a momentum microscope\cite{kspace}.  This photoelectron analyzer detects the complete angular distribution of the photoelectrons for a selected kinetic energy ($E_{kin}$), as exemplarily shown in the insets of Fig 2b for two selected values of $E_{kin}$.   Varying $E_{kin}$ allows to record a 3D data set of the ARPES intensity as function of electron momentum parallel to the surface. In this way, energy distribution curves for all high symmetry directions are recorded simultaneously (see Methods for further informations). The OPO laser system is used as excitation source for 2PPE. By varying the  photon energy (between 4.13\,eV and 4.43\,eV) and the light polarization (between $s$ and $p$), we can easily assign the features in the ARPES spectra to either {\it occupied} or {\it unoccupied} electronic states (i.e.\ states {\it below} or {\it above} $E_F$) and determine their surface- or bulk-related character. Crucially, using photon energies above 4\,eV gives us access to the still unexplored region of the Brilloiun zone  far away from the $\bar{\Gamma}$ point, where we expect a strong deviation of the dispersion of the Au(111) SSs from trivial Rashba SSs.\\
By a careful analysis of the 2PPE ARPES data (see Supplementary Information) we can identify three dominant  contributions to the spectra: the occupied and unoccupied part of the SSs, an unoccupied image potential resonance, and bulk states.  
The dispersion of the SSs and of the image potential resonance are plotted in Fig 2b with black and blue circles, respectively. From previous ARPES \cite{Reinert:2001} and STS  \cite{Schouteden:2009}  studies it is well known that the occupied Shockley SSs can be described by a quasi-free electron parabola with an effective mass $m_{eff}$ in the range of $0.25\,m_e$\cite{Reinert:2001} to $0.37\,m_e$ \cite{Woodruff:1986}. The significantly smaller effective mass compared to the free electron mass ($m_e$) is due to an intrinsic coupling of the SSs with bulk states \cite{Unal:2011}. Our analysis reveals an effective mass of $0.30\,m_e$ (red solid line in Fig 2b), in good agreement with previous studies. The unoccupied image potential resonance also follows a free electron like behavior as it is well known from literature \cite{Woodruff:1986}.
The unoccupied part of the SSs, on the other hand, show a clear deviation from the trivial free electron like behavior. In particular, in the energy range from 2.0\,eV to 3.0\,eV above $E_F$, the unoccupied SSs are found at larger $k_{||}$-values than expected for a free electron like behavior with $m_{eff} =0.30\,m_e$. For the Shockley SS of the Cu(111) surface, \"{U}nal {\it et al}.\ found a similar deviation  below the Cu L-bulk band edge \cite{Unal:2011} and explained it as the result of hybridization of the SS with the Cu bulk bands. The strength of the hybridization, i.e., the deviation of the experimental dispersion from the free electron like behavior, increases as the SSs approach the L-band edge. 
Crucially, for intermediate state energies larger than $3.0$\,eV  the dispersion of the SSs in our data changes again and the SSs disperse faster to larger momentum values with increasing energy above $E_F$. To our knowledge, such a strong deviation from the free electron like behavior was not observed yet for SSs on fcc(111) noble metal surfaces.  This observation can be explained by assuming that the SSs disperses into the bulk bands for intermediate state energies larger than $3.0$\,eV in order to connect the valence and conduction bands. This behavior, together with the dispersion of the SSs, is in full agreement with the calculations in Fig 2a, pointing to the topological nature of the SSs. \\
To conclusively demonstrate the topological nature of the SSs we now turn to the \textit{ab-initio} calculations. Here, we increase the strength of SOC artificially and realize an indirect energy gap in bulk Au, for example, when the SOC strength is 350\%  of the normal one. The motivation for this approach is that topological nature of the SSs has been long-neglected plausibly due to the lack of an energy gap in gold. Thus, we  design here a real gap to demonstrate these SSs derived from topology. In the bulk, the band structures with 100\% SOC strength and 350\% SOC strength are adiabatically connected to each other, exhibiting the same topology with $\nu_0=1$ (see Supplementary Information). On the surface, an energy gap opens in the band structure for the 350\%  SOC case (see Fig. 2b). Inside this gap, a pair of gapless SSs appears, with one branch merging into conduction bands and the other branch into valence bands. The existence of a single Fermi surface~\cite{fu2007b} between $\bar{\Gamma}$ and $\bar{M}$ points provides unambiguous evidence of TSSs. Because SSs of the  100\% SOC case are adiabatically connected to TSSs of the  350\% SOC case, we can conclude that the normal SSs on the Au(111) surface are also TSSs. 
In addition, Ag and Cu exhibit bulk and surface band structures that are very similar to those of Au (see Supplemental Information)  and equivalent in topology. Therefore, we conclude that Au, Ag, and Cu are all topological metals wherein TSSs exist on the surface. We note that the spin splitting observed on Cu(111) is surprising larger~\cite{Tamai2013} than expected from the Rashba effect based on the weak SOC of Cu~\cite{Reinert2003}, indicating the topological origin of SSs.
We note that TSSs should also exist on other facets of these noble metals owing to the nontrivial $Z_2$ index of the bulk. This is consistent with surface states, e.g., on (110) and (001) surfaces, reported in the literature~\cite[and references therein]{Goldmann1985,Smith1985}.
The topological energy gap above the Fermi energy is similar to the case of a recently discovered oxide TI, BaBiO$_3$~\cite{Yan2013}. It is also reminisces another topological metal, Sb. Bulk Sb is a semimetal with a direct energy gap and a nontrivial band structure~\cite{fu2007a}. As a consequence, TSSs have been observed on special facets such as the Sb (111) and (110) surfaces in experiments~\cite{Sugawara2006,hsieh2009bisb,Seo2010,Bianchi2012}. 

After we clarify the topological feature of Au, Ag, and Cu, we further generalize the same idea to other FCC noble metals: Pt and Pd. These two metals show very similar bulk band structure to Au. The main difference is that their Fermi energies are lower than that of Au because Pt and Pd have one fewer valence electron  than Au. Therefore, TSSs also exist on both Pt and Pd (111) surfaces, as shown in Fig. 3. Compared to those of the Au surface, these TSSs shift downward toward the bulk bands in energy and but still lie above the Fermi energy for both Pt and Pd surfaces. For the Pt (111) surface, the Dirac point of TSSs is found to slightly merge into the bulk bands and forms a surface resonance. In the literature, these empty SSs have been observed and also interpreted as Shockley SSs with Rashba-splitting  in photoemission and scanning tunneling spectroscopy (STS) for both Pt~\cite{Roos1995,Wiebe2005,Bendounan2011} and Pd~\cite{Hulbert1986}. We note that the positions of TSSs in these reports~\cite{Roos1995,Wiebe2005,Hulbert1986} are consistent with our results. For example, STS revealed the unoccupied SSs of the Pt (111) surface at $\bar{\Gamma}$ with strong SOC splitting above the Fermi energy~\cite{Wiebe2005}.  {\color{black} Here we call these SSs on noble metal surfaces as topologically derived surface states (TDSSs), to distinguish them from TSSs on a real insulator.}

Both TDSSs and Shockley SSs are in-gap states and originate from the inversion of two bulk bands with different symmetries. The TI requires an odd number of band inversions at TRIM in the Brillouin zone, and TDSSs are protected by  time-reversal symmetry (TRS). In contrast, Shockley states also need band inversions, but they are not limited to finite positions of the zone and to a finite number. Hence, Shockley states may exist more commonly than TDSSs and lack the protection by TRS.
{\color{black} The robustness of TSSs, which refers to their existence inside the inverted energy gap in the energy spectra,
 is protected by the band topology of the bulk. }
Such  topological protection was indeed observed for TDSSs of noble metals in previous experiments. These states remain robust for example under the adsorption of alkali metals~\cite{Lindgren1979,Kevan1986,Tang1993,Sandl1994}, guest noble metals~\cite{Bendounan2011}, rare gases~\cite{Forster2003,Forster2004}, and CO and oxygen~\cite{Lindgren1979,Paul1981,Ozawa2014} and even against surface reconstruction~\cite{LaShell1996,Reinert2004}. 
In contrast to trivial SSs such as dangling bond states, they usually shift in energy rather than get eliminated by adsorbates or reconstruction~\cite{Roos1995}.
{\color{black}  But the topological protection does not necessarily mean the robustness against any types of
electron scattering for example by surface defects.}
Because the Dirac cone is heavily deformed, with both  upper and lower cones as well as bulk bands crossing the Fermi energy, TDSSs can be backscattered on the metal surface, which is different from those of a TI with an energy gap
~\cite{hasan2010,qi2011RMP}. This {\it weakness} allows versatile manipulation of these SSs by  quantum confinement (e.g., in quantum corrals).  Therefore, we conclude that TDSSs on noble metal surfaces are more stable than trivial SSs in energy spectra due to topological protection,  but more fragile than TSSs of a real TI regarding to surface scattering. In addition, the Dirac point is expected to open an energy gap by breaking TRS, e.g., by depositing magnetic impurities on the surface, as previously observed in magnetic TIs~\cite{Chen2010c,Xu2012}.

Because Au, Pt, and Pd are important catalysts, the existence of TDSSs provokes an interesting question on the role of topological states on the catalysis process. 
TDSSs determine the surface charge density at the Fermi energy and thus sensitively influence surface-related processes. 
 Although the underlying mechanism calls for further investigation,
TSSs supply an interdisciplinary platform upon which chemistry meets modern condensed-matter physics.

\paragraph{Methods}
The \textit{ab-initio} calculations have been performed within the framework of density-functional theory(DFT) with the generalized gradient approximation~\cite{perdew1996}.  We employed the Vienna ab initio simulation
package with a plane wave basis~\cite{kresse1996}. The core electrons were represented by
the projector-augmented-wave potential. The bulk band structure of Au was projected to Au-$sp$ orbitals in Fig.1c using Wannier functions~\cite{Mostofi2008}. The surface band structures were calculated on a slab model that includes thirty-three atomic layers using DFT.\\
The ARPES experiments were performed in an ultrahigh vacuum (UHV) setup with a base pressure of $10^{-11}$\,mBar, equipped with a momentum microscope \cite{kspace} for the  detection of the complete angular distribution of the photoelectrons as a function of their kinetic energy.  The Au(111) surface was prepared {\it in situ} by repeated cycles of Ar$^{+}$ ion sputtering at 2\,kV and annealing to 570\,K of a mechanically polished (111) crystal.
For the 2PPE experiments, we used a commercial OPO laser system (Inspire OPO, Spectra Physics) with a pulse width of 150\,fs. The laser was focused on the sample with a spot size of a few micrometers at an incidence angle of $65^{\circ}$ with respect to the sample surface.   The photon energy was tuned between $4.13$\,eV and $4.43$\,eV, while the light polarization was changed between $s$ and $p$ using a $\lambda/2$ plate.

\begin{addendum}
\item [Acknowledgements] We thank Prof. S. S. P. Parkin at IBM Almaden Research Center San Jose and Prof. S.-C. Zhang at Stanford University for fruitful discussions. B.Y. and C.F. acknowledge financial support from the ERC Advanced Grant (291472) and computing time at HLRN Berlin/Hannover (Germany).


\item[Competing Interests] The authors declare that they have no competing financial interests.

 \item[Correspondence] Correspondence and requests for materials
should be addressed to B.Y.~(email: yan@cpfs.mpg.de).
\end{addendum}


\clearpage
\begin{figure}
   \centering
   \includegraphics[width=1\linewidth]{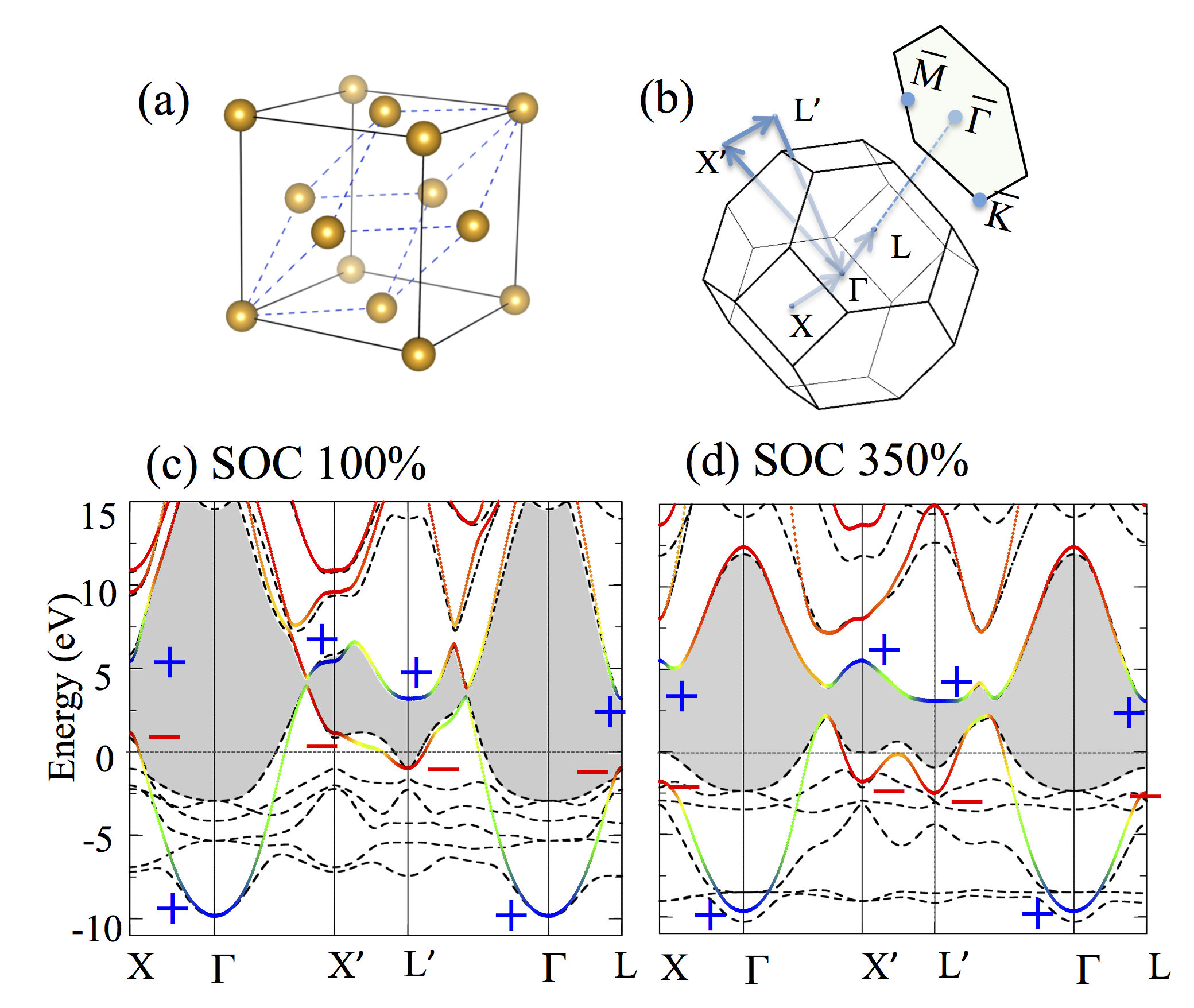} 
\end{figure}
\noindent {\bf Fig. 1.} 
Crystal structure and band structures of bulk gold. (a) Face-centered cubic lattice structure. The dashed lines indicate the primitive unit cell that includes only one atom. (b) The first Brillouin zone of the bulk and the surface Brillouin zone projected onto the (111) surface. The band structure of the bulk with (c) normal SOC strength (100\%) and (d) enhanced SOC to 350\%. Band structures are 
 plotted along the high symmetry lines $X$ (0.5 0.5 0) $-$  $\Gamma$ (0 0 0) $-$ $X^{\prime}$ (0.5 $-$0.5 0) $-$ $\Gamma$ $-$ $L^{\prime}$ (0.5 $-$0.5 0.5) $-$ $\Gamma$ $-$ $L$ (0.5 0.5 0.5), in which coordinates are in units of reciprocal lattice vectors.  Dashed black lines are \textit{ab-initio} results while color lines are projection to Au-$sp$ states through Wannier functions.
The color gradient from blue to red represents varying contributions from $s$ to $p$ states.
The parity eigenvalues of $s$-type (blue, ``$+$'') and $p$-type (red, ``$-$'') wave functions  are labeled in the band structure. 
The gray shadow region indicates the topological energy gap that are induced by the $sp$ band inversion.
The Fermi energy is shifted to zero and indicated by the horizontal dotted line.
The $s$ and $p$ bands get inverted in order at $X$ ($X^\prime$) and $L$ ($L^\prime$) points while they remain in the normal order only at the $\Gamma$ point, which makes the band structure topologically nontrivial with $Z_2$ index $\nu_0 = 1$. 

\begin{figure}
   \centering
   \includegraphics[width=1\linewidth]{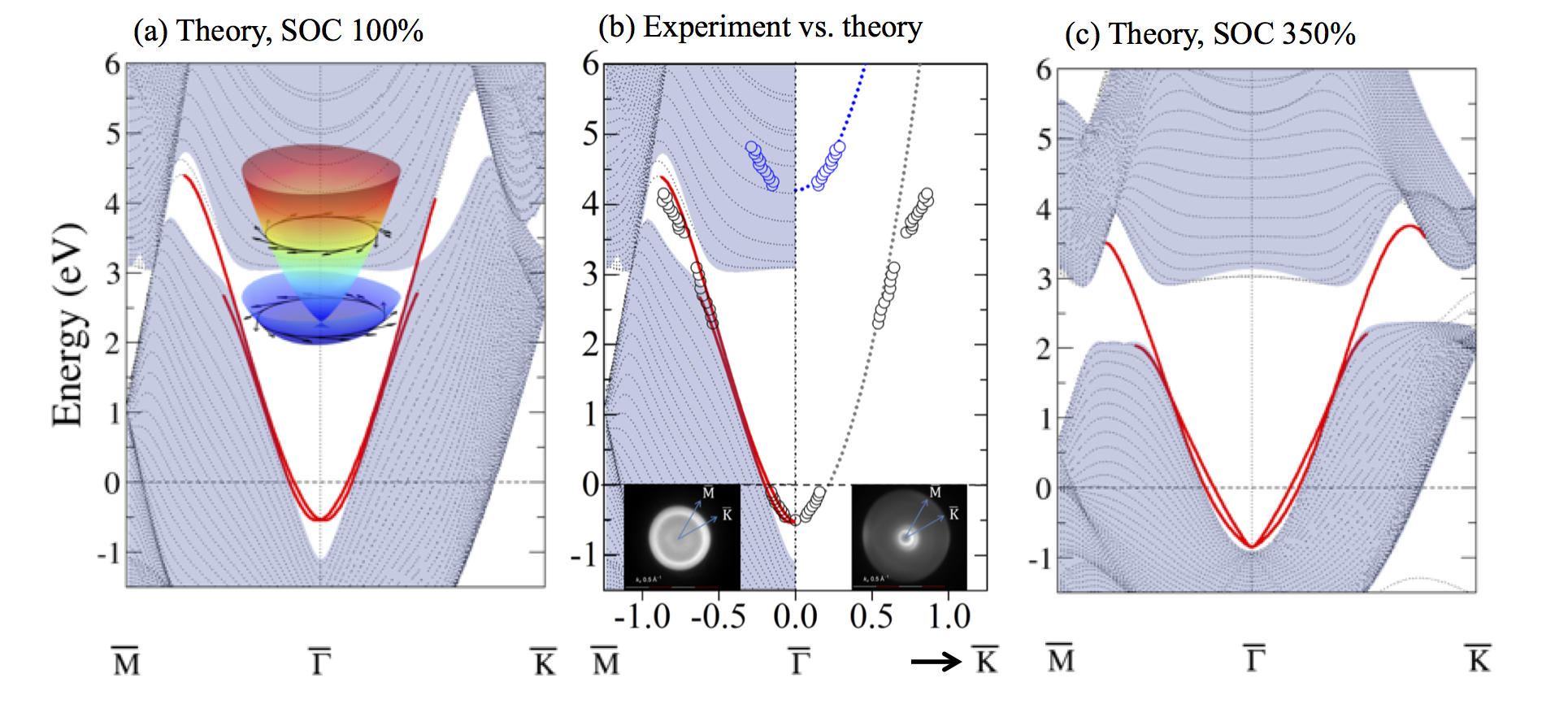} 
\end{figure}
\noindent {\bf Fig. 2.} 
Panels (a) and (c) show the surface band structure of the Au(111) surface calculated for a normal SOC strength (100\%) and an enhanced SOC strength  of 350\%. The red lines indicate the topological protected surface states (TSSs). The inset of (a) illustrates the three-dimensional plot of near Dirac point with helical spin texture. The bulk projections are indicated by the light blue shadow.  In panel (b), the dispersion of the Au(111) surface state (black points) is superimposed to the theoretical prediction for normal SOC (red curve). The experimental data points have been extracted from the  2PPE ARPES measurements performed with the OPO laser system combined with the momentum microscope. This experimental set-up allows recording maps with the angular distribution of the photoelectrons at konstant kinetic energy ($E_{kin}$) values for all available kinetic energies. The two maps corresponding to $E_{kin}=2.6\,eV$ and $E_{kin}=4.0\,eV$ are shown exemplarily in the inset of (b). (For more details see supplementary materials.) The blue data points illustrate the dispersion of the image potential resonance.

\begin{figure}
   \centering
   \includegraphics[width=1\linewidth]{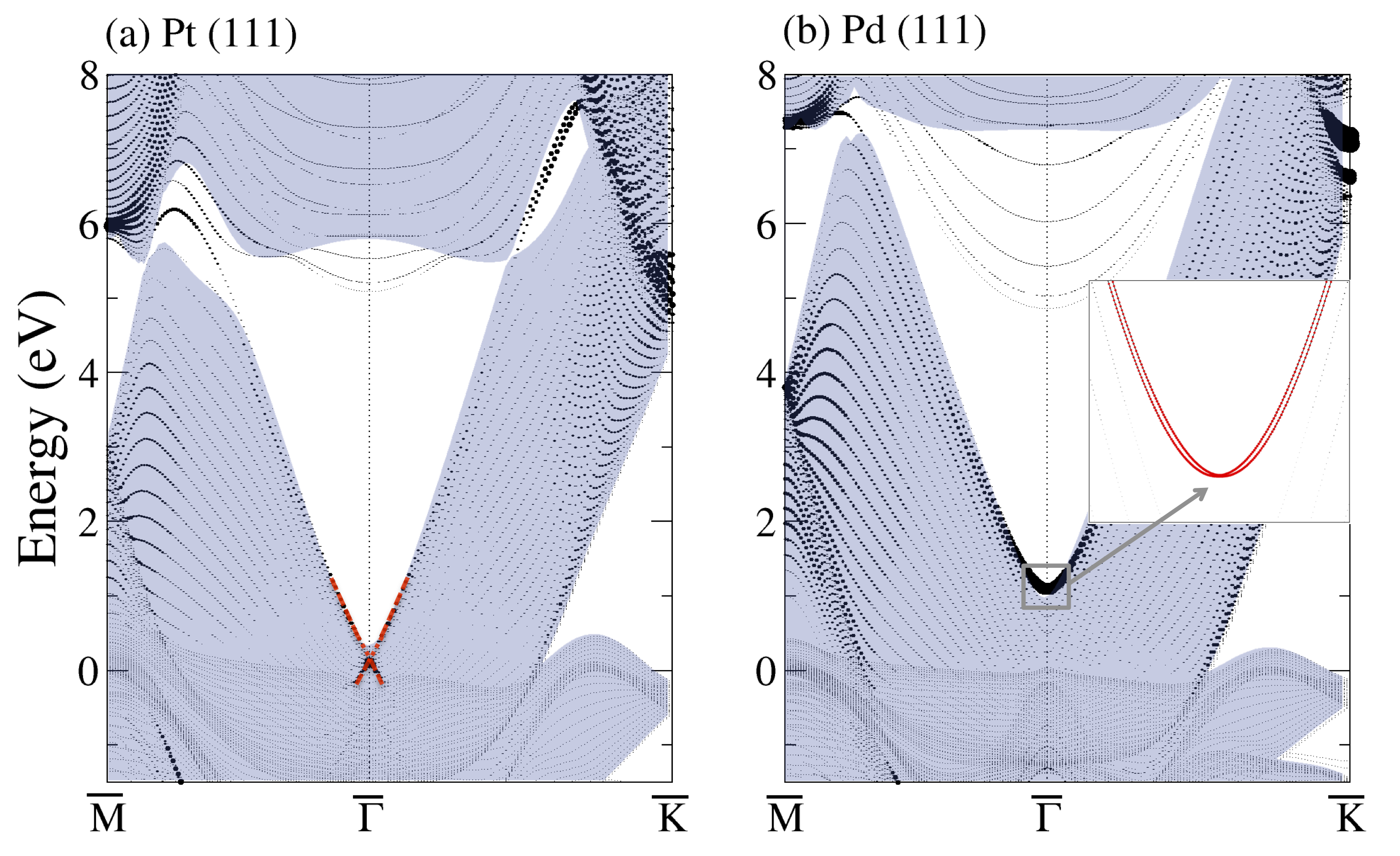} 
\end{figure}
\noindent {\bf Fig. 3.} 
Surface band structure of (a) Pt (111) and (b) Pd (111) surfaces with normal SOC strength. The size of the filled circles represents the amplitude of projection to the $sp$ orbitals of the top two surface atomic layers. Red lines highlight the topological surface states to guide the eyes in (a). The topological surface states are shown enlarged by filled red circles in (b). The Fermi energy is shifted to zero and topological surface states remain unoccupied above the Fermi energy.

\end{document}